\documentclass[
    ,final            
  ]
  {aipproc}

\layoutstyle{8x11double}

\begin{document}

\title{What Makes the Crab Pulsar Shine?}


\author{J. A. Eilek}{
  address={New Mexico Tech, Socorro NM, USA}
}

\author{T. H. Hankins}{
   address={New Mexico Tech, Socorro NM, USA}
}


\begin{abstract}
  Our high time resolution observations of individual pulses from the
  Crab pulsar show that the main pulse and interpulse differ in
  temporal behavior, spectral behavior, polarization and dispersion.
  The main pulse properties are consistent with one current model of
  pulsar radio emission, namely, soliton collapse in strong plasma
  turbulence.  The high-frequency interpulse is quite another story.
  Its dynamic spectrum cannot easily be explained by any current
  emission model; its excess dispersion must come from propagation
  through the star's magnetosphere. We suspect the high-frequency
  interpulse does not follow the ``standard model'', but rather comes
  from some unexpected region within the star's magnetosphere.
  Similar observations of other pulsars will reveal whether the radio
  emission mechanisms operating in the Crab pulsar are unique to that
  star, or can  be identified in the general population.
\end{abstract}


\maketitle


 
How do pulsars make radio emission?  What are the physical conditions
in the magnetosphere that make the pulsar ``shine'' in the radio?  Is
there one answer to this question, or many?  What observable
quantities can identify the emission mechanism, or distinguish between
different theories of the emission mechanism?  In order to answer
these questions, we have developed specialized data acquisition
systems and used them to observe individual radio pulses at the
highest possible time resolution.  Our goal is to find ways to use our
data to confront competing theories of pulsar radio emission.

Up to now we have concentrated on the Crab pulsar, because it
occasionally emits very bright pulses which are ideally suited to our
data acquisition systems.  The mean profile of the Crab pulsar is
dominated by a Main Pulse (MP) and an Interpulse (IP), which can be
identified from low radio frequencies (below $\sim$ 100 MHz) up to optical,
X-ray and $\gamma$-ray bands ({\it e.g.}, \cite{MH1996}).  The
similarity of the mean profile across this broad frequency range
suggests that both radio and high-energy emission come from the same
regions in the magnetosphere.  

In traditional pulsar models, the MP and the IP are thought to come from
plasma, at relatively low altitudes, moving out along open field lines
from the star's two magnetic poles.  Other geometrical models have
been also been suggested, such as emission from higher-altitude
``outer gaps'' or ``caustics'' \cite{DyksHard, CR}
above each magnetic pole, or
broad outflows over one magnetic pole \cite{Manch}.  While these models
differ in geometry, they all agree that physical conditions should be
similar in the regions which emit the MP and IP.  Because similar
physics should lead to similar emission processes, these models
suggest that the MP and IP should be the same in their observable
quantities (such as spectrum, time signature, or dispersion).  We were
--- and remain --- quite surprised that this turns out {\it not} to be
the case in the Crab pulsar.

\begin{figure}[ht]
\vspace{-1.5in}
\rotatebox{-90}{
  \includegraphics[trim = 40 0 60 0, width=0.6\textwidth,clip]{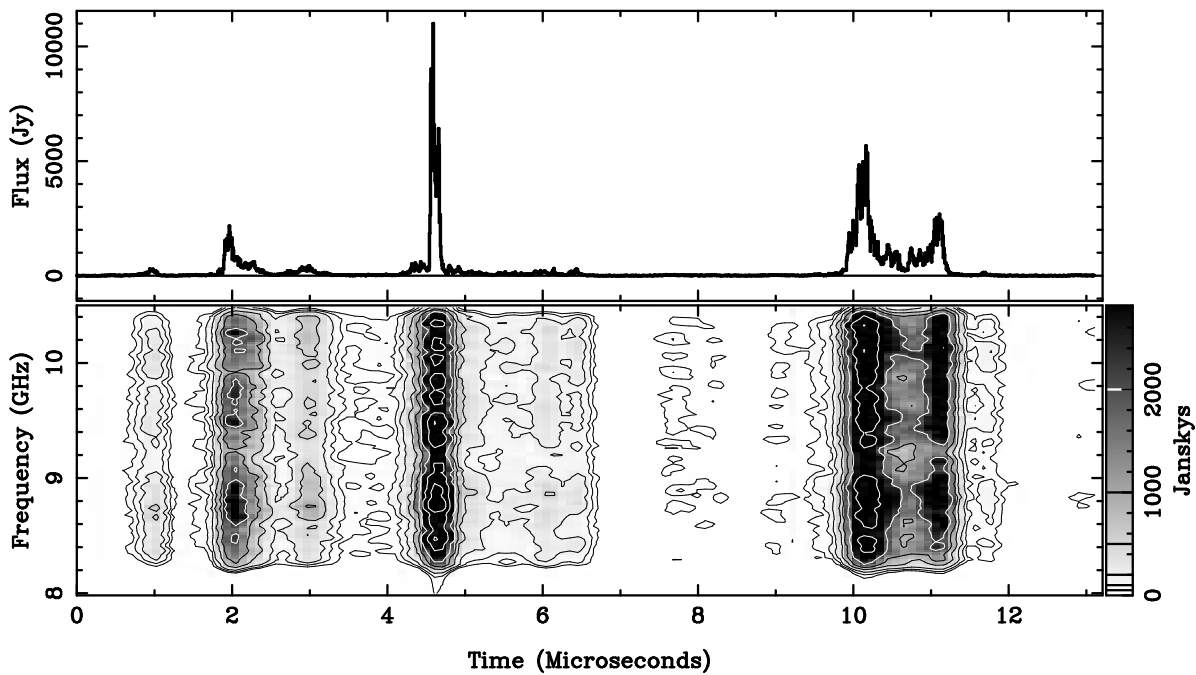}}
\vspace{-2.5in}
\caption{An example of a Main Pulse, observed with 2.2-GHz bandwidth
  at 9 GHz, and coherently dedispersed \cite{HKWE}. The pulse seen in
  total intensity (upper panel, plotted with total intensity time
  resolution 6.4 ns) contains several short-lived microbursts. The
  dynamic spectrum (lower panel, plotted with resolution 102 ns and
  19.5 MHz) shows that the microburst emission spans the full receiver
  bandwidth.  In a few MPs individual, short-lived nanoshots are
  sparse enough in time to be separately identified (shown in
  \cite{HKWE, HE2007}); these examples reveal the dynamic spectrum of
  the nanoshots is relatively narrow. }
\label{MPfig}
\end{figure}

\begin{figure}[ht]
\rotatebox{-90}{
  \includegraphics[trim = 40 0 60 0, width=0.6\textwidth,clip]{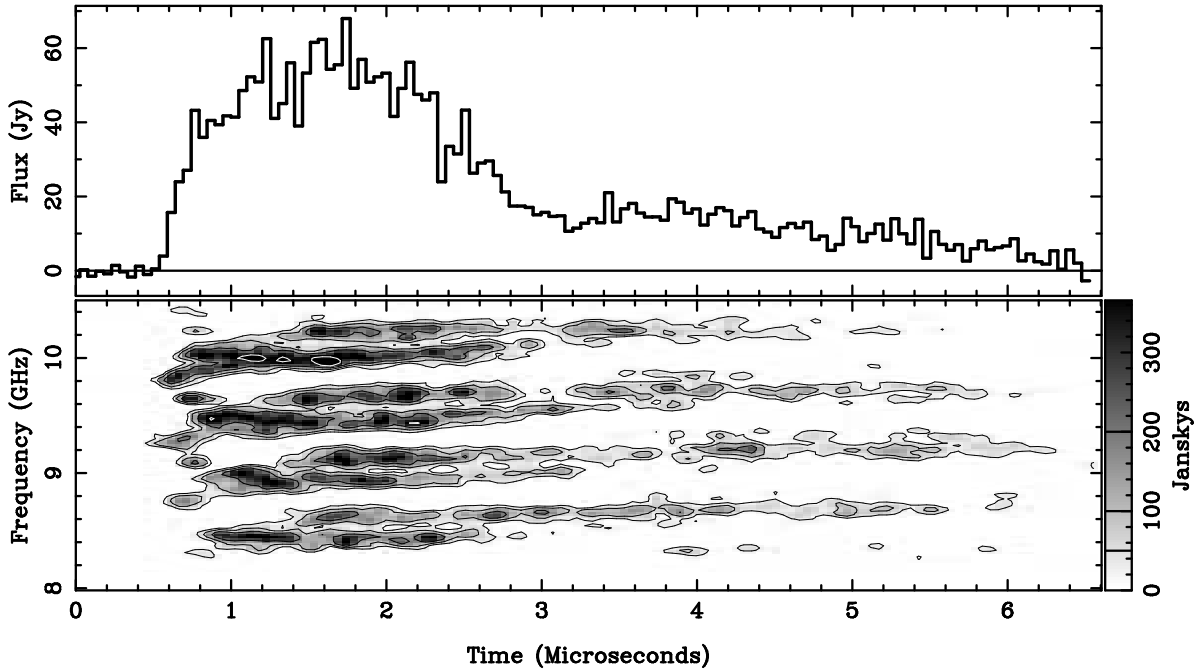}}
\caption{An example of an Interpulse, observed with 2.2 GHz bandwidth
  at 9 GHz, and coherently dedispersed.  The IP seen in total
  intensity (upper panel) typically contains 1 or 2 sub-bursts; thus
  it has a simpler time signature than the MP (as in the example of Figure
  \ref{MPfig}).  The regular {\it emission bands} in the dynamic
  spectrum (lower panel) are not due to instrumental or interstellar
  effects, but are characteristic to the emission physics of the IP.
  Note that the secondary burst, seen in total intensity, coincides
  with the appearance of new band sets in the dynamic spectrum.
  Plotted with total intensity time resolution 51.2 ns, and dynamic
  spectrum resolution 104 ns and 19.5 MHz. }
\label{IPfig}
\end{figure}

In this paper we compare and contrast properties of the MP
and IP, and consider what our results tell us about pulsar radio
emission physics. We illustrate our discussion with two example
pulses, in Figures \ref{MPfig} and \ref{IPfig}; more details and
examples can be found in \cite{HKWE} and \cite{HE2007}.

\section{Our observations}

Between 1994 and 2002 our group observed strong pulses from the
Crab  pulsar between 1 and 5 GHz at the VLA and Arecibo.  Our Arecibo
observations were designed with 2-ns time resolution, in order to test
competing models of the radio emission mechanism.  We initially
concentrated on the MP, because it is stronger than the IP below
$\sim$5 GHz in the star's mean profile, and also because giant pulses
are more common at the phase of the MP at these frequencies
\cite{Cordes2004}.  Our results, and our data acquisition system, are
described in \cite{HKWE}.

To follow up on these results, we  extended our data
acquisition system to higher time resolution.  We went to higher
frequencies (5 to 10 GHz) in order to take advantage of the 2.2-GHz
bandwidths (and corresponding sub-ns time resolution) available at
Arecibo. We recorded individual IPs as well as MPs, because at these
frequencies strong pulses are much more common at the phase of the IP
\cite{Cordes2004}.  These new observations, carried out between 2003
and 2006, are reported in \cite{HE2007}.  We were astonished to find
that IPs are very different from MPs at these frequencies. The IP
differs from the MP in polarization, time signature, spectrum and
dispersion, as discussed below.

All of our IP observations were taken betweeen 5 and 10 GHz.
Technical limitations, as well as the scarcity of strong IPs at lower
frequencies, kept us from observing the IP below 4 GHz.  We are
therefore describing the ``high-frequency IP'', which occurs at a
slighly earlier rotation phase from the ``regular'' IP (as seen at lower
radio frequencies as well as in  high-energy bands; {\em e.g.}, 
\cite{MH1996}).  This phase offset suggests that the high-frequency IP
may not be related at all to the regular IP; it may come from a very
different part of the magnetosphere.

\section{Different time signatures}

Most MPs contain several short-lived microbursts, as illustrated in
Figure \ref{MPfig}.  At 1.4 GHz the microbursts are typically $\sim$
3-30 $\mu$s long, with some tendency for more powerful bursts to be
shorter-lived \cite{Jared}.  At this frequency the microbursts have a
fast-rise, slow-decay shape.  Below $\sim$1 GHz, the pulse width is
determined by interstellar scattering.  At higher frequencies, however,
the burst width is larger than $\nu^{-4}$ extrapolation predicts  \cite{Kern}, 
and the burst width 
can change by more than a factor of ten from burst to
burst \cite{Jared}.  Thus, the microburst width above $\sim$1 GHz is not due 
to interstellar effects;  we 
are observing the temporal profile of the burst as it leaves the
star. At higher frequencies, the microbursts are shorter;  their width
scales approximately $\propto \nu^{-2}$ \cite{Kern}.  They lose their
fast-rise, slow-decay profile, and become more symmetric in time, as
illustrated in Figure \ref{MPfig}. 

 An occasional MP can 
be resolved into well-separated, short-lived ``nanoshots''.  In
\cite{HKWE} we reported 5-GHz nanoshots shorter than 2 ns; in our
later work \cite{HE2007} we found similar nanoshots at 7 and 9 GHz,
some of which remain unresolved at 0.4 ns.  We suspect that all MP
microbursts are ``clouds'' of overlapping nanoshots;  only
rarely are the nanoshots sparse enough to be identified individually.

IP emission is more continuous in time than MP emission.  It usually
begins with a rapid onset, followed by slower decay, as seen in Figure
\ref{IPfig}. Although a second burst can often be identified,
overlapping in time with the first burst, IP emission is not broken up
into the short-lived microbursts that characterize the MP. The IP
lasts $\sim$1-3 $\mu$s at 9 GHz (as measured by the equivalent width
of the autocorrelation function), and somewhat longer at lower
frequencies, approximately consistent with the $\nu^{-2}$ behavior of
MP microbursts.

\section{Different spectral signatures}

MP microbursts tend to be broadband, emitting across our
full 2.5-GHz bandwidth, as Figure \ref{MPfig} illustrates.  The
emission is sometimes, but not always, weaker towards the high end of
the band;  this is  probably consistent with the known steep spectrum of the
Crab pulsar. However, when we captured individual (sparse) nanoshots
between 6 and 10 GHz, we found that their spectra are relatively narrow-band, 
$\delta \nu / \nu \sim 0.1-0.2$ \cite{HE2007}.  If all MP microbursts are indeed
collections of nanoshots, then their broad-band spectrum does not reflect
the fundamental emission process, but comes from a 
composite of overlapping nanoshots.  The steep
radio spectrum of the MP is due either to nanoshots at higher
frequencies being fewer, or fainter (or both).

The dynamic spectrum of the IP is dramatically different. As Figure
\ref{IPfig} illustrates, the IP spectrum contains regular
\emph{emission bands}, which we have detected from 5 to 10 GHz.  All
the bands within one set turn on at the same time (to within $\sim$100
$\mu$s).  A single IP usually contains more than one band set;
additional sets turn on $\sim$1-2 ns later in the pulse, often
shifted to a slightly higher frequency. Band sets which start later in
a given IP can often be identified with a new ``burst'' in the
total intensity profile.  The center frequency of each band in a band
set usually remains constant, but sometimes drifts slightly upwards
during the few-$\mu$s lifetime of the band set.

Although at first glance the bands appear regularly spaced, in reality
they are \emph{proportionately spaced}:  the frequency separation
between two adjacent bands is $6\%$ of their mean frequency.
We have never seen  band sets that do not span our full 
observed bandwidth (2.2 GHz);  we therefore suspect the 
bands extend from 5 to at least 10.5 GHz in a single IP.  Because the
high-frequency IP does not continue below $\sim 4$ GHz in the mean
profile, we speculate that the emission bands we observe do not
continue below that frequency.

\section{Different polarization}

The mean-profile MP is weakly polarized at 1.4 and 5 GHz (typically
20\%; \cite{MH1999}).  Individual nanoshots in a single MP, however,
can be strongly polarized, but the sense of the polarization can vary
from one nanoshot to the next \cite{HKWE}.  This tells us that the
intrinsic MP emission process is highly polarized, but no local
``memory'' in the emission region retains the sense of polarization
from one nanoshot to the next.

Once again the IP behaves differently.  It is strongly linearly
polarized (at least 50\%) at 5 and 8 GHz in the mean profile
\cite{MH1999}.  Thus, the intrinsic IP emission process is highly
polarized, and some local ``memory'' here  does retain the
sense of polarization from one IP to the next.

\section{two types of radio emission}

The dramatic temporal, spectral and polarization differences between
the MP and IP point to the existence of two quite different mechanisms
for coherent radio emission operating in this star.

The time and spectral signatures of the MP appear to be consistent
with existing models of pulsar radio emission. In \cite{HKWE} we
compared time signature predictions of three classes of radio emission
mechanisms then in the literature: coherent charge bunches, stimulated
emission/masers, and soliton collapse in strong plasma turbulence.  We
argued that only one of these three --- soliton collapse, as modelled
numerically by \cite{JCW} --- has a characteristic time which is short
enough to be consistent with the nanoshots we observed at 5 GHz.  The
narrow-band nature of individual nanoshots, which we found in
\cite{HE2007}, is also consistent with \cite{JCW}.  We therefore
propose that the individual, short-lived nanoshots which comprise
 MP emission are
produced by soliton collapse in strong plasma turbulence.

The unusual dynamic spectrum of the IP suggests that it involves quite
different emission physics from MP emission, and probably requires some
new thinking. We are not aware of any models of pulsar radio emission
which predict such regular emission bands over such a broad frequency
range.  In particular, plasma resonant emission --- which has been a
popular model for pulsar radio emission --- tends to be narrow-band,
centered on a frequency determined by plasma density and/or magnetic
field in the emission region.

As a naive example, if this were coherent plasma emission, the IP
emission region would have to be unusually stratified, containing 15
density steps, each 3\% higher than its neighbor, in order to produce
the proportionately spaced emission bands we see between from 5 to 10 GHz. In
order to turn on within 100 ns of each other, these stratifications
would have to be co-located to within 30 m. We find such structures
unlikely. Alternatively, the IP emission might come from a
high-altitude cyclotron resonance, as suggested by \cite{Maxim}.
Again,  stringent  relations between the plasma density, particle
energy and viewing angle are required in order to reproduce the full
range of proportional band spacing; it is not clear that such
conditions will arise naturally in the star's  magnetosphere. 

In \cite{HE2007} we speculated that the bands may be an interference
phenomenon.  We tend to like this idea, but emphasize that many
details must be worked out before this speculation can be described as
a ``model''.  In particular, we need an intrinsically broad-band emission
mechanism (such as a double layer), which coexists with regular,
small-scale plasma structures (to provide the interference).
How to make and maintain such structures is far from obvious.

\section{Higher interpulse dispersion}

The IP is more dispersed than the MP.  The dispersion of individual
MPs does not vary significantly from pulse to pulse (to less than
$\sim .001$pc-cm$^{-3}$), and is approximately consistent with the
monthly values monitored at Jodrell Bank (based on mean profiles at
lower frequency: \cite{Lyne}). The dispersion of individual IPs,
however, varies substantially from pulse to pulse during a single run,
by $ \sim.01$pc-cm$^{-3}$.  On average IPs are more dispersed than MPs
by about this same amount.  Our methods and results regarding MP/IP
dispersion will be reported in \cite{Crossley}.

 Although pulsar dispersion is traditionally assumed to be due to the
 signal's passage through the interstellar medium (ISM), this picture
 cannot explain why IPs have higher dispersion than MPs observed a few
 minutes later.  It follows that the IP signal must gain the extra
 dispersion {\it during its passage through the star's magnetosphere};
the extra dispersion induced by that passage can vary
 significantly within a few minutes.  Our data also suggest that the
 IP dispersion law has a flatter frequency dependence than the
 standard cold-plasma dispersion law which describes ISM dispersion  ---
 but further observations are needed to confirm this result.

\section{Summary:  the Crab pulsar}

We have identified two different types of coherent radio emission from
the Crab pulsar, one associated with the MP, the other associated with
the high-frequency IP.  Two different radiation mechanisms seem to be
be operating within the star's magnetosphere.  In
addition, the higher dispersion of the high-frequency IP suggests the
signal has passed through an unusually large plasma column before
leaving the pulsar.

Conventional wisdom ascribes the MP to emission from the open field
line region above one of the star's magnetic axes.  If this is the
case, the high-frequency IP is probably not simply radiation from the
other magnetic pole. It is more likely to come from some unexpected
part of the magnetosphere; its phase offset, relative to the regular
IP \cite{MH1996}, corroborates this idea. The suggestion
from \cite{Maxim} that the high-frequency IP 
is emitted within the closed field line
region may be on the right track.

\section{What about  other pulsars?} 

Our results apply to strong pulses from the Crab pulsar;  do they have any
relevance to other pulsars?  Is the Crab pulsar a useful  example of
pulsar radio emission physics? 

Some might argue that the Crab pulsar is so unusual that our results
should not be applied to other stars.  It is, of course,
possible to argue that the high-flux ``giant'' pulses which we capture
are not typical of more common ``weak'' pulses which are usually
recorded from this star \cite{Lundgren} (although the observational
issues are not yet resolved; {\em e.g.}, \cite{PopStap}).  It is also
true, of course, that in many ways the Crab pulsar is not typical of
the general pulsar population. It is a young pulsar, with unusually
strong high-energy pulsed emission.  Its mean profile is unusual in
both the number and phase location of emission components, and its
unusual polarization sweep \cite{MH1999}.  It shows more ``giant''
pulses than most pulsars do (although new work on intermittent pulsars
suggests pulsar duty cycles are not yet understood {\em e.g.},
\cite{PW}).

We nonetheless  think our results are broadly applicable.  The
phenonema that make the Crab pulsar unusual are {\it macroscopic}, in
that they depend on dynamics of the magnetospheric plasma.  On the other
hand, coherent radio emission is a {\it microscopic}
process. Plasma dynamics provide a source of free energy;  if plasma conditions
are favorable, locally coherent charge motions in the plasma 
convert that free energy to intense radio emission.  More than one
macroscopic situation  can lead to the same microscopic
physics, and thus the same radio emission.

As a concrete example, we know beam instabilities can drive the strong
plasma turbulence which we believe is responsible for coherent radio
emission from the MP.  Beams may come from relative streaming of
electrons and positrons in an imperfectly shielded electric field, for
instance in the open field line region close to the star's magnetic
poles (as in the traditional view of pulsars).  Alternatively, we know
that local reconnection events can drive beams (as happens in solar
flares).  In pulsars such events may be driven by high-altitude plasma
shear in the dynamic upper magnetosphere ({\em e.g.}, \cite{Maxim03,
Spitkv}).  The observable properties of 
coherent radio emission from both scenarios could
be very much the same.

We therefore argue that our results on the Crab pulsar may well be
relevant to other pulsars; but this speculation must be backed up by
observations.  The next step, therefore,  is to look at other pulsars
with similar time and frequency resolution.  To do this will require the
sensitivity of the largest radio telescopes and the widest practical
bandwidths.


\begin{theacknowledgments}
  The results we report here would not have been possible without the
  long and hard work contributed by our students past and present,
  especially J. Crossley, D. Moffett, J. Kern and J. Sheckard.  This
  work was partially supported by NSF grants AST-0139641 and
  AST-0607492.  NRAO and AUI operate the VLA, as NAIC and Cornell operate
  Arecibo,  under cooperative agreements with NSF.
\end{theacknowledgments}

\end{document}